\batchmode
\makeatletter
\def\input@path{{\string"C:/Users/Anderson/Dropbox/Pesquisa/artigos/Experimental determination of OAM in complex shaped vortices/submissao/\string"/}}
\makeatother
\documentclass[twocolumn,american,aps, prl, showpacs, reprint]{revtex4}
\usepackage[T1]{fontenc}
\usepackage[latin9]{inputenc}
\usepackage{amsmath}
\usepackage{amssymb}
\usepackage{graphicx}
\usepackage{esint}

\makeatletter
\@ifundefined{textcolor}{}
{%
 \definecolor{BLACK}{gray}{0}
 \definecolor{WHITE}{gray}{1}
 \definecolor{RED}{rgb}{1,0,0}
 \definecolor{GREEN}{rgb}{0,1,0}
 \definecolor{BLUE}{rgb}{0,0,1}
 \definecolor{CYAN}{cmyk}{1,0,0,0}
 \definecolor{MAGENTA}{cmyk}{0,1,0,0}
 \definecolor{YELLOW}{cmyk}{0,0,1,0}
}

\makeatother

\usepackage{babel}
\begin{document}

\title{Distinguishing orbital angular momenta and topological charge in
optical vortex beams}

\author{Anderson M. Amaral}

\email{anderson.amaral@outlook.com}

\affiliation{Departamento de Física, Universidade Federal de Pernambuco, 50670-901
Recife, PE, Brazil}

\author{Edilson L. Falcão-Filho}

\affiliation{Departamento de Física, Universidade Federal de Pernambuco, 50670-901
Recife, PE, Brazil}

\author{Cid B. de Araújo}

\email{cid@df.ufpe.br}

\affiliation{Departamento de Física, Universidade Federal de Pernambuco, 50670-901
Recife, PE, Brazil}
\begin{abstract}
In this work we discuss how the classical orbital angular momentum
(OAM) and topological charge (TC) of optical beams with arbitrary
spatial phase profiles are related to the local winding density. An
analysis for optical vortices (OV) with non-cylindrical symmetry is
presented and it is experimentally shown for the first time that OAM
and TC may have different values. The new approach also provides a
systematic way to determine the uncertainties in measurements of TC
and OAM of arbitrary OV.

\end{abstract}


\maketitle
Optical vortices (OV) have been extensively studied since the seminal
work by Allen et al. \citep{allenorbital1992} and are applied in
topics as diverse as classical and quantum communications \citep{WillnerTerabitTelecom2012,ZeillingerOAMentanglement2001,plickquantum2013},
optical tweezers \citep{Ritsh-Marte2008,DholakiaPerfectVortexDynamics2013}
and plasmonics \citep{Lee2010,rurycoherent2013,toyodaOVtwistednanostructures2012,gorodetskiOAMfromPlasmon2013,brasselettopological2013}.
However, there are subtleties in characterizing such optical beams
that are not usually remarked. Canonical OV, as associated to Bessel
or Laguerre-Gauss beams, carry well defined mean values of orbital
angular momentum (OAM) and topological charge (TC). In these cases,
the OAM per photon and the TC have the same value. However, by analyzing
these quantities for non-canonical OV it can be seen that they represent
distinct quantities. A proper understanding of non-canonical OV is
of great interest because they extend the current applications of
OV. For example, it is possible to control transverse forces in optical
tweezers \citep{Ritsh-Marte2008}, or increase the excitation efficiency
of surface plasmon modes \citep{brasselettopological2013}. In the
present work we explicitly distinguish classical and modal OAM and
TC for arbitrarily shaped OV beams. This is fundamental to avoid mistakes
when analyzing the experimental consequences of more general beams.

Diffractive and interferometric techniques as \citep{soskintopological1997,berkhoutmultipinhole2008,Hickmann2010}
are sensitive to the phase profile, hence they measure the total topological
charge (TC) of a beam. The quantum OAM distribution may be obtained
via diffractive elements \citep{PadgettOAMSorting2010,BoydOAMSorting2013}
or modal decomposition \citep{LofflerOAMsidebands2012,ForbesModeDecomposition2013}.
The classical OAM of a light beam may be determined by measuring the
electric field amplitude and phase, as in \citep{Padgettfractionavortex2004},
or also via modal decomposition \citep{ForbesModeDecomposition2013}.

We consider a scalar OV beam under the paraxial approximation. In
cylindrical coordinates $\mathbf{r}=\mathbf{r}\left(\rho,\phi,z\right)$
a linearly polarized and monochromatic field in vacuum may be represented
by the following vector potential 
\begin{equation}
\mathbf{A}\left(\mathbf{r},t\right)=\hat{\epsilon}\left(\frac{2\mu_{0}}{\omega k}P_{0}\right)^{\frac{1}{2}}\mathcal{A}\left(\mathbf{r}\right)\exp i\left[\chi\left(\mathbf{r}\right)+kz-\omega t\right],\label{eq:field}
\end{equation}
where $P_{0}$ is the optical power, $\mu_{0}$ is the vacuum permeability,
$\omega,k$ are respectively the angular frequency and wave number
of light. The remaining terms are the beam phase profile $\chi\left(\mathbf{r}\right)$
and $\mathcal{A}\left(\mathbf{r}\right)$ is the vector potential
amplitude envelope, normalized such that $\int d^{2}r\left|\mathcal{A}\left(\mathbf{r}\right)\right|^{2}=1$.
The total TC, $Q_{T}$, contained inside a contour $C$ of radius
$c$, on the $\rho\phi$ plane, is given by \citep{Nakahara,Berry2004}
\begin{align}
Q_{T} & =\frac{1}{2\pi}\oint_{C}d\mathbf{x}\cdot\nabla\chi\left(\mathbf{r}\right)=\frac{1}{2\pi}\int_{0}^{2\pi}w\left(\rho=c,\phi\right)d\phi,\label{eq:TC}
\end{align}
where the local winding density (LWD) $w\left(\mathbf{r}\right)$
is a quantity that gives the local effect of the TC, is defined by
\begin{equation}
w\left(\mathbf{r}\right)=\frac{\partial\chi}{\partial\phi}\left(\mathbf{r}\right).\label{eq:winding}
\end{equation}

The total TC gives the number of times that the beam phase pass through
the interval $\left[0,2\pi\right]$ following the curve $C$. For
a well-behaved contour, $Q_{T}$ is an integer even if $\chi\left(\mathbf{r}\right)$
is discontinuous \citep{Berry2004}. 

On the other hand, the classical OAM density along the propagation
direction $\hat{z}$ is $L_{z}=\frac{\epsilon_{0}\omega}{2}\mathfrak{Re}\left\{ \mathbf{A}\left(\mathbf{r},t\right)\left(-i\frac{\partial}{\partial\phi}\right)\mathbf{A}^{*}\left(\mathbf{r},t\right)\right\} $
\citep{Berryparaxialbeams1998}, and it may be shown by direct substitution
of eq. \eqref{eq:field} that 
\begin{equation}
L_{z}=\frac{P_{0}}{\omega c}\left[\mathcal{A}^{*}\left(\mathbf{r}\right)\frac{\partial\chi}{\partial\phi}\mathcal{A}\left(\mathbf{r}\right)\right].\label{eq:Lz}
\end{equation}

Since $P_{0}=N\hbar\omega$, where $N$ is the number of photons impinging
on the plane $\rho\phi$ per second carrying energy $\hbar\omega$,
the local OAM value per photon at a given position is \citep{PadgettMathieu2002}
$\hbar w=\hbar\partial\chi/\partial\phi$. So, the LWD gives the local
OAM per photon. 

We remark that although the intensity profile of a beam is related
to its TC distribution \citep{Berry2004,Grier2003,Shaping_Optical_Beams},
the intensity profile carries no information about the topological
or OAM properties of a beam \citep{pugatchtopological2007}.

Since the product $\mathcal{A}^{*}\left(\mathbf{r}\right)\mathcal{A}\left(\mathbf{r}\right)$
gives the probability of finding a photon at a given point, the average
classical OAM per photon may be determined from eq. \eqref{eq:Lz}
as \citep{PadgettMathieu2002} 
\begin{equation}
\hbar\left\langle l\right\rangle _{class.}=\hbar\int d^{2}r\mathcal{A}^{*}\left(\mathbf{r}\right)\frac{\partial\chi}{\partial\phi}\mathcal{A}\left(\mathbf{r}\right).\label{eq:average_Lz}
\end{equation}

A comparison between eqs. \eqref{eq:TC} and \eqref{eq:average_Lz}
shows that only in very specific situations $Q_{T}=\left\langle l\right\rangle _{class.}$.
An immediate result is that measuring $Q_{T}$ one does not necessarily
have information about $\left\langle l\right\rangle _{class.}$ and
vice-versa. However, both quantities are related to the LWD which
can be obtained from $\chi\left(\mathbf{r}\right)$. Therefore we
emphasize that it is important to determine the LWD for the characterization
of the classical OAM and TC in OV beams.

In a quantum description of OAM, it can be shown that eq. \eqref{eq:average_Lz}
gives the correct average OAM %
\footnote{see Supplemental Material at <URL>.%
}. Thus, if $\left|p,m\right\rangle $ represents the radial and azimuthal
quantum numbers $p$ and $m$, respectively, it can be seen that the
average OAM is given by 
\begin{equation}
\hbar\left\langle l\right\rangle _{quant.}=\hbar\sum_{p,m}m\left|\left\langle \mathcal{A}|p,m\right\rangle \right|^{2},\label{eq:Average_Lz_Quantum}
\end{equation}
while a comparison with the classical expression, eq. \eqref{eq:average_Lz},
shows that, as expected by the correspondence principle, $\hbar\left\langle l\right\rangle _{quant.}=\hbar\left\langle l\right\rangle _{class.}$
and then {[}34{]}

\begin{equation}
\hbar\int d^{2}r\mathfrak{Re}\left\{ \left\langle \mathcal{A}|\mathbf{r}\right\rangle \hat{L}_{z}\left\langle \mathbf{r}|\mathcal{A}\right\rangle \right\} =\hbar\left\langle l\right\rangle _{class.}.
\end{equation}

\begin{figure}[h]
\begin{centering}
\includegraphics[width=1\columnwidth]{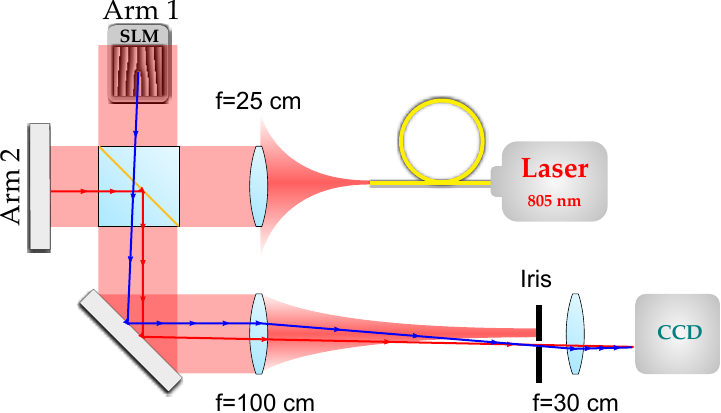}
\par\end{centering}

\caption{Experimental setup (not to scale). The output of a fiber coupled laser
diode emitting at $805$ nm is collimated with a lens with long focal
distance ($f=25$ cm), producing a nearly plane wave. The collimated
light goes to a Michelson interferometer in which the arm 1 contains
a SLM (Hamamatsu - LCOS X10468-02). The arm 2 provides the plane wave
reference, and it is blocked when beam intensity measurements are
performed. The reference (red) and modulated (blue) beams have a small
relative angle and are spatially filtered and then imaged on a CCD
camera (Coherent - Lasercam HR) positioned at the SLM image plane.
\label{fig:Experimental-setup}}
\end{figure}

The setup shown in Fig. \ref{fig:Experimental-setup} allows a full
characterization of a linearly polarized electric field by measuring
its amplitude and phase. It consists of a Michelson interferometer
in which Arm 1 contains a spatial light modulator (SLM). When Arm
2 is blocked, only the intensity profile will be detected by the CCD,
otherwise an interference pattern will be detected. From the interference
pattern, $\chi\left(\mathbf{r}\right)$ may be retrieved using Fourier
transforms \citep{Takeda-Fourier-transform1982}. To obtain a better
signal/noise ratio, we averaged the phase for 20 applied constant
phase offsets on the SLM \citep{Padgettfractionavortex2004}. To compute
the azimuthal derivative in the LWD, we used a Fourier spectral method
with a smoothing gaussian filter \citep{ahnertnumerical2007}. To
determine $w\left(\mathbf{r}\right)$, we used the following identity
$w\left(\mathbf{r}\right)=\partial\chi/\partial\phi=e^{-i\chi}\left[-i\left(x\frac{\partial}{\partial y}-y\frac{\partial}{\partial x}\right)\right]e^{i\chi}.$

\begin{figure}[h]
\begin{centering}
\includegraphics[width=1\columnwidth]{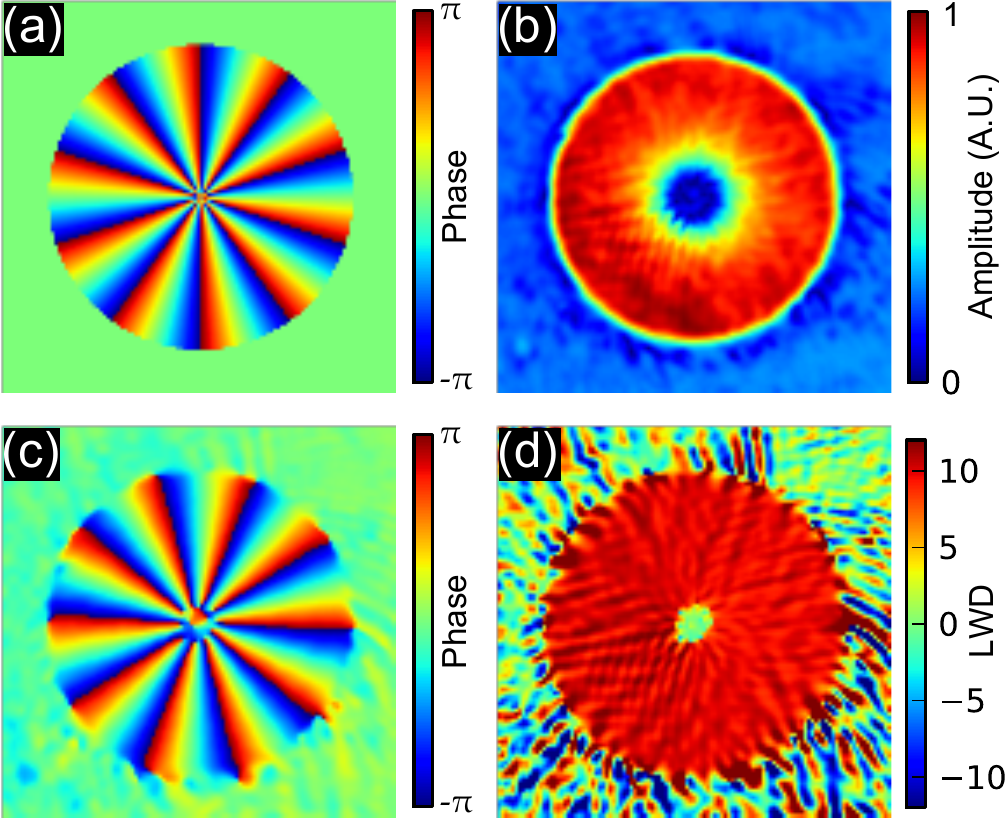}
\par\end{centering}

\caption{(color online) Typical profiles for a vortex with azimuthal phase
$\chi_{sig}=10\phi$. (a) Phase pattern applied to the SLM (without
carrier). (b) and (c) contains, respectively, the experimentally measured
amplitude and phase profiles, averaged in 20 samples. Notice that
the experimental phase profile (d) agree well with the phase applied
at the SLM (a). (d) Spatial profile of the LWD. \label{fig:Typical-properties}}
\end{figure}

The first set of measurements was obtained with beams having a linear
azimuthal phase dependence $\chi_{sig}=\alpha\phi$, for integer $\alpha$.
For all measurements, the SLM phase profile was composed of $\chi_{sig}$,
a circular aperture with fixed radius and a carrier wave. A typical
$\chi_{sig}$ is shown in Fig. \ref{fig:Typical-properties} (a) for
$\alpha=10$. The measured amplitude and phase profiles are shown,
respectively in Figs. \ref{fig:Typical-properties} (b-c) and the
corresponding LWD is shown in Fig. \ref{fig:Typical-properties} (d).
Notice that, as expected, $w\left(\mathbf{r}\right)$ is well defined
along the beam profile, except where the intensity is very small and
the phase is not well retrieved. Outside the beam intense region there
is a background due to light diffracted from the SLM which adds systematic
phase and LWD shifts.

For a quantitative description of the classical OAM and the TC, we
notice that eq. \eqref{eq:TC} can be considered as an average of
the LWD over a narrow ring and \eqref{eq:average_Lz} is an average
of the LWD weighted by $\left|\mathcal{A}\right|^{2}$. So, weighting
the LWD with $K^{\mbox{TC}}=1$ over a ring ($K^{\mbox{OAM}}=\left|\mathcal{A}\right|^{2}$
over the beam) one may build histograms representing the probability
$P^{\mbox{TC}}$ ($P^{\mbox{OAM}}$) of finding a given value $w'$
of LWD (OAM) in a narrow range, $\epsilon\ll1$, and whose average
is $Q_{T}$ ($\left\langle l\right\rangle _{class.}$) via 
\begin{align}
P^{\beta}\left(w'\right) & =\frac{\int d^{2}r\,\theta\left[\epsilon-\left|w\left(\mathbf{r}\right)-w'\right|\right]K^{\beta}}{\int d^{2}r\, K^{\beta}},\label{eq:ProbTCOAM}
\end{align}
where $\theta$ is the step function and $\beta$ corresponds to TC
or OAM.

\begin{figure}[h]
\begin{centering}
\includegraphics[width=1\columnwidth]{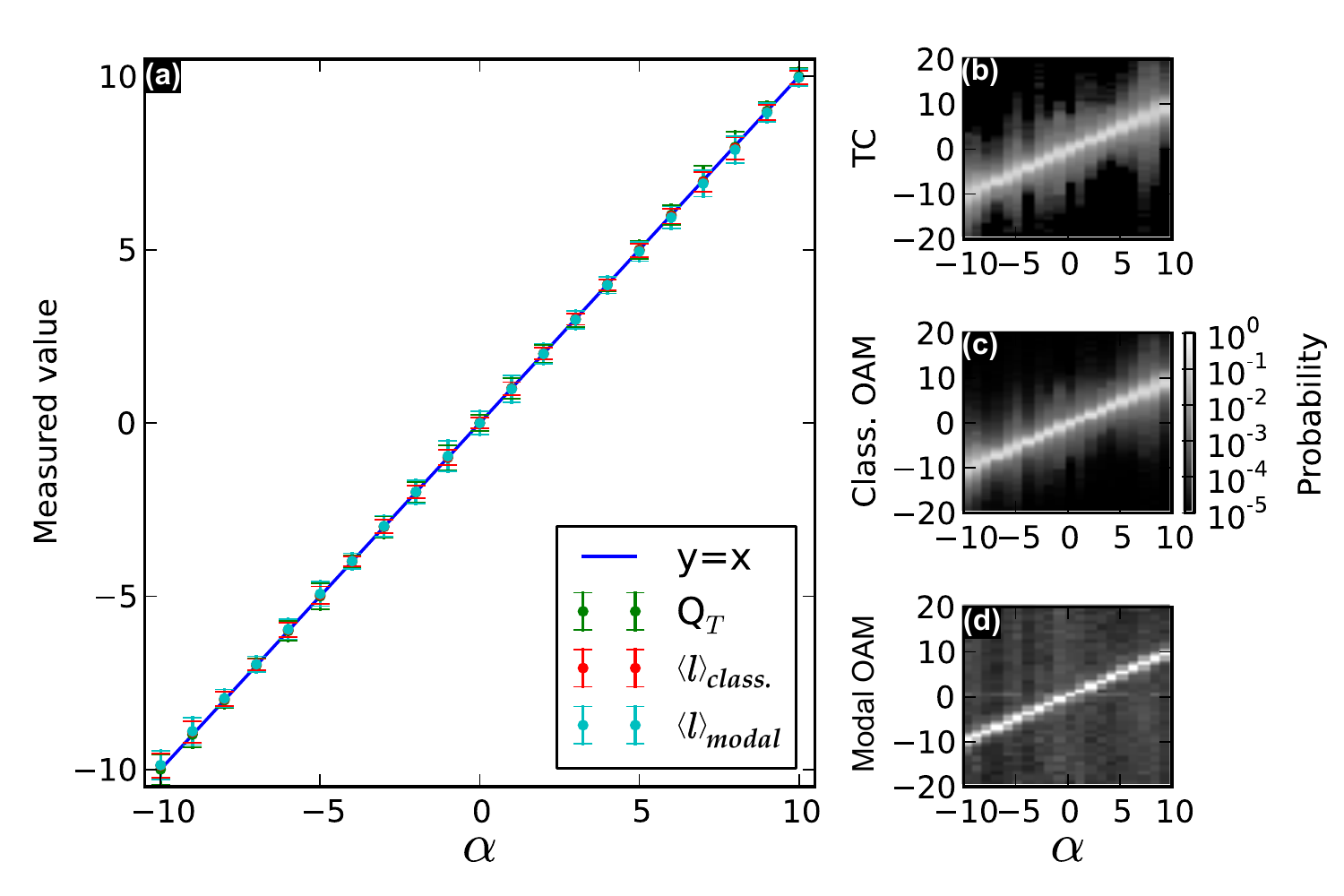}
\par\end{centering}

\caption{(color online) (a) Superimposed plots of measured TC, classical OAM
via \eqref{eq:average_Lz} and modal OAM via full mode decomposition,
eq. \eqref{eq:Average_Lz_Quantum}. No distinction is observed for
all quantities. (b), (c) and (d) are, respectively, the histograms
associated with TC, classical OAM and modal OAM. The curve $y=x$
is added as a visual guide.\label{fig:Pure azimuthal phase}}
\end{figure}

The TC and classical OAM histograms, as defined above, allow quantitative
determination of the uncertainties of the measured quantities. To
the best of our knowledge, no previous classical characterization
of OAM and TC was able to determine these uncertainties. We also produced
histograms for the modal OAM distribution, which is related to the
quantum OAM, by expanding the field in a basis of Bessel functions.
It can be seen in Fig. \ref{fig:Pure azimuthal phase} (a) that $\left\langle Q_{T}\right\rangle $,
$\left\langle l\right\rangle _{class.}$ and $\left\langle l\right\rangle _{modal}$
are similar for integer $\alpha$. The associated histograms for each
$\alpha$ can be seen in Figs. \ref{fig:Pure azimuthal phase} (b-d).
Except for the distinction of quasi-continuous distribution in Figs.
\ref{fig:Pure azimuthal phase} (b-c) for TC and classical OAM and
the discrete distribution for modal OAM, no major differences are
observed. 

\begin{figure}
\begin{centering}
\includegraphics[width=1\columnwidth]{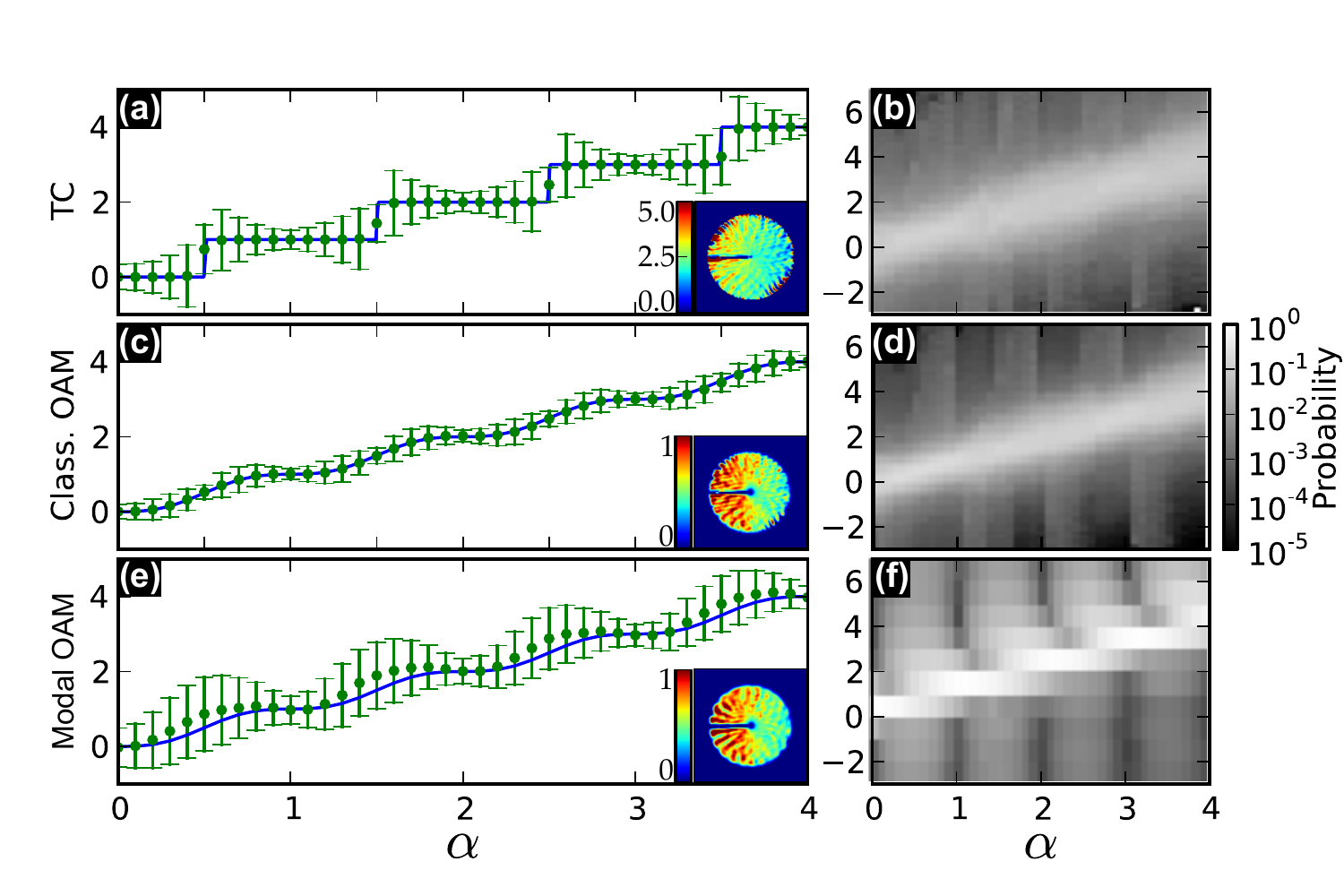}
\par\end{centering}

\caption{(color online) Experimental determination of TC (a), classical OAM
via eq. \eqref{eq:average_Lz} (c) and modal OAM via eq. \eqref{eq:Average_Lz_Quantum}
(e). The insets contain the spatial profiles of the LWD (a), and the
classical OAM density profile obtained from classical and modal expressions
(c, e) for $\alpha=2.5$. The theoretical predictions are represented
by continuous lines. (b), (d) and (f) correspond, respectively, to
the histograms for the TC, classical OAM and modal OAM. \label{fig:Fractional_vortex}}
\end{figure}

On the other hand, TC and OAM for fractional values of $\alpha$ behave
differently from integer ones. As suggested in \citep{Berry2004},
$Q_{T}$ is the nearest integer to $\alpha$. Simultaneously, it
can be shown that the OAM for fractional $\alpha$ is $\left\langle l\right\rangle =\alpha-\sin\left(2\pi\alpha\right)/2\pi$
\citep{Padgettfractionavortex2004,BarnettQuantumFractionalOV2007}.
The experimental histograms for OAM and TC are shown in Figs. \ref{fig:Fractional_vortex}
(a,c,e). In Fig. \ref{fig:Fractional_vortex} (a) it is shown how
$\left\langle Q_{T}\right\rangle $ varies with $\alpha$. In Figs.
\ref{fig:Fractional_vortex} (c,e) it can be seen that $\left\langle l\right\rangle $
follows smoothly the theoretical prediction. The insets in Fig. \ref{fig:Fractional_vortex}
(a, c, e) exhibit respectively, the LWD profile (eq. \eqref{eq:winding}),
and local probability densities for classical OAM (integrand of eq.
\eqref{eq:average_Lz}) and modal OAM (integrand of eq. \eqref{eq:Average_Lz_Quantum}
obtained from the Bessel expansion). We remark that modal OAM histogram
also behaves as is theoretically expected \citep{BarnettQuantumFractionalOV2007}.

\begin{figure}
\begin{centering}
\includegraphics[width=1\columnwidth]{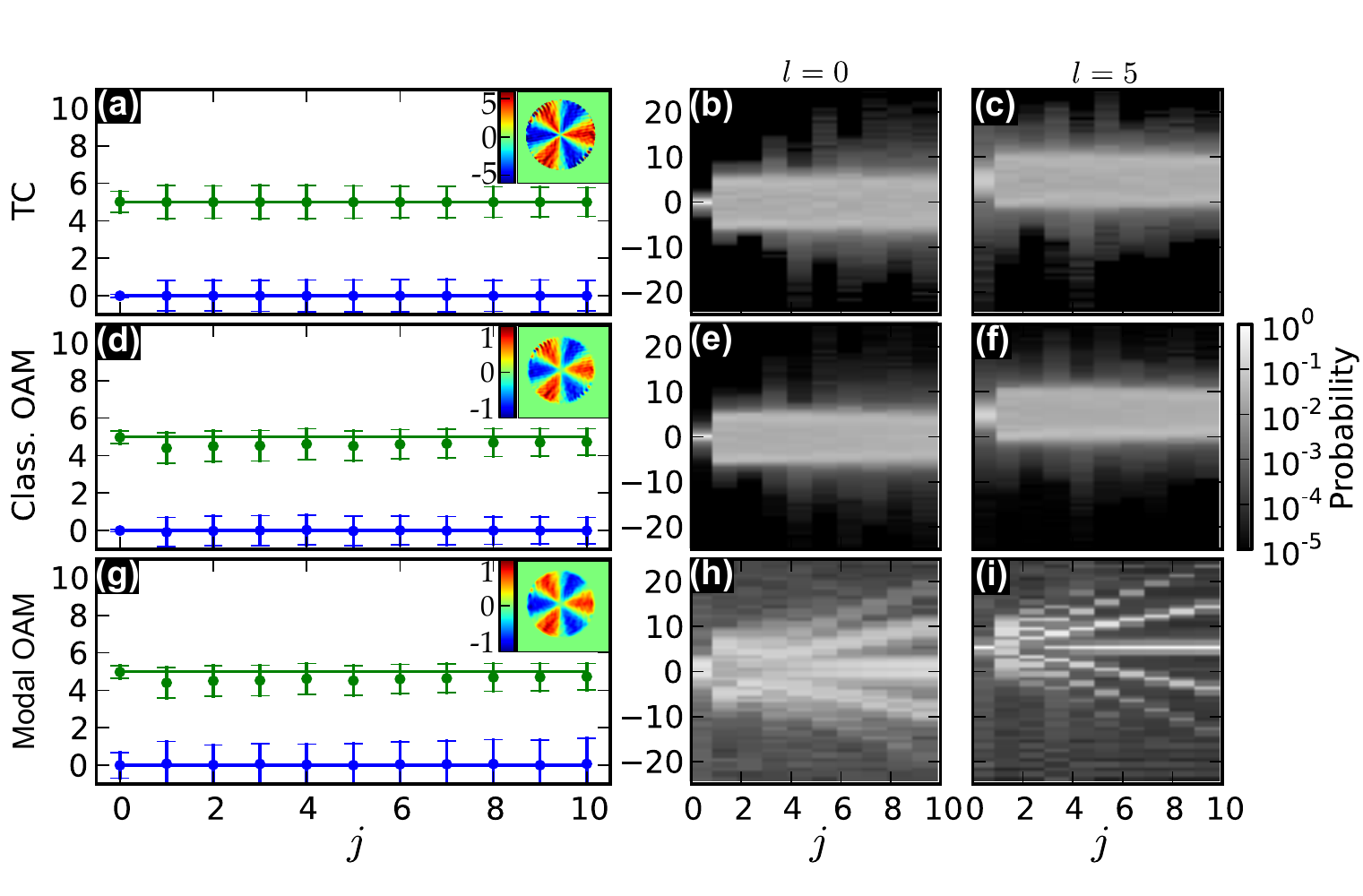}
\par\end{centering}

\caption{(color online) (a, d, g) TC, classical OAM and modal OAM for Lissajous
OV with $a=5$ and $l=0$ (blue) or $l=5$ (green). Solid lines represent
the theoretically expected values. The insets contain the spatial
profiles of the LWD (a), and the classical OAM density profile obtained
from classical and modal expressions (d, g) for $l=0,\, j=3$. (b,
e, h) Histograms of TC, classical OAM and modal OAM for $l=0$. (c,
f, i) Histograms of TC, classical OAM and modal OAM for $l=5$.\label{fig:shaped-1}}
\end{figure}

In Fig. \ref{fig:shaped-1} we consider Lissajous-shaped OV, in which
the OV cores form Lissajous patterns \citep{Grier2003}. The phase
and LWD profiles of these OV may be written respectively as$\chi_{\mbox{Lissajous}}=l\phi+a\sin\left(j\phi\right)/j$
and $w_{\mbox{Lissajous}}=l+a\cos(j\phi)$, with $j\neq0$.

Since the oscillatory term averages to zero, it is expected that $\left\langle Q_{T}\right\rangle =\left\langle l\right\rangle =l$,
and this can be observed in Figs. \ref{fig:shaped-1} (a, d, g). Notice
that, the histograms for $l=5$ in Fig. \ref{fig:shaped-1} (c, f,
i) are shifted with respect to those with $l=0$ in Fig. \ref{fig:shaped-1}
(b, e, h). An interesting feature of Lissajous OV is in the comparison
between the classical and modal OAM. Approximating the beam by a top-hat,
the classical OAM histogram is not sensitive to the number of $w$
oscillations. Therefore it depends only on $l$ and $a$ as is observed
for $j\neq0$ in Figs. \ref{fig:shaped-1} (e, f). Meanwhile, the
modal OAM depends on $j$, as seen in Figs. \ref{fig:shaped-1} (h,
i) and from the probability of obtaining the OAM eigenvalue $l'$
in a Lissajous OV in terms of Bessel functions $P\left(l'\right)=\left|J_{\frac{l'-l}{j}}\left(\frac{a}{j}\right)\right|^{2}$
\citep{Gradshteyn}.

Notice that the different $j$ dependence between the classical and
modal OAM of Lissajous OV may be used to distinguish classical and
quantum OAM transfer. 

\begin{figure}
\begin{centering}
\includegraphics[width=1\columnwidth]{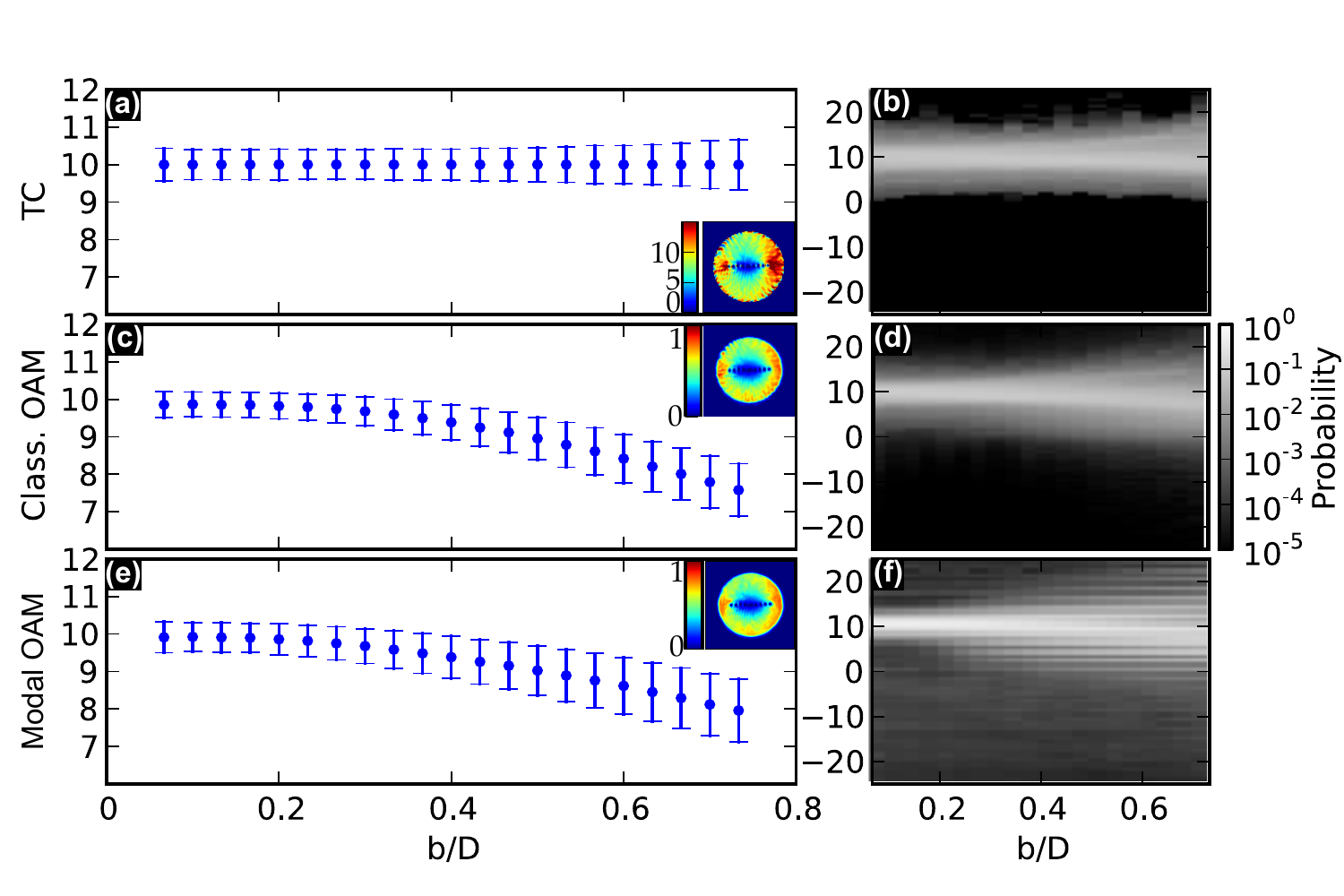}
\par\end{centering}

\caption{(color online) Line of equally charged OV for $D=3.3$ mm. (a) Total
TC, (c,e) classical and modal OAM. On insets are represented the spatial
profiles of the LWD (a), and the classical OAM density profile obtained
from classical and quantum expressions (c, e) for $b/D=0.6$. In (b,d,f)
are represented the histograms respective to (a,c,e). \label{fig:LineOV}}
\end{figure}

Finally we consider a linear distribution of OV, each having a unitary
TC \citep{Shaping_Optical_Beams}. The OV are uniformly spaced over
a line of length $b$ and inside a circular intensity envelope of
diameter $D$. It can be shown that using such distribution of TCs
one may shape the OV core for small TC separation \citep{Shaping_Optical_Beams}.
The experimental results are shown in Fig. \ref{fig:LineOV}. As a
first remark, it may be noticed that, since the total TC is the sum
of the individual TCs, it remains constant for all $b/D$ values in
Fig. \ref{fig:LineOV} (a). The TC histogram in Fig. \ref{fig:LineOV}
(b), varies little with $b/D$ and essentially becomes slightly broader
for larger $b/D$. A different behavior is observed for the OAM. For
large $b/D$ values, the average OAM is reduced because the phase
due to each OV is compensated between equally charged vortices and
such regions become more illuminated for larger TC separations. This
fact can also be seen from a full description of such TC distribution
\citep{Indebetouw1993}. The classical and modal OAM histograms of
Figs. \ref{fig:LineOV} (d, f) have similar profiles, the modal one
being grainier. The similarity in this non-cyllindrical OV configuration
is interesting because it corroborates the correctness of our interpretation
of eq. \eqref{eq:ProbTCOAM} as a classical probability of finding
a given OAM in a light beam.

In summary, we remark that by analysis of the LWD one may obtain spatial
information about the TC and the classical OAM content of a light
beam. The characterization presented is applicable to arbitrarily
shaped OV and clarifies that TC and OAM are usually different quantities.
Such distinction is important because it raises fundamental questions.
For example, are the increased coherence storage times for OV reported
in \citep{pugatchtopological2007} due to TC or OAM? Does the excitation
of surface plasmons in \citep{Lee2010,toyodaOVtwistednanostructures2012,gorodetskiOAMfromPlasmon2013,brasselettopological2013}
depends on mode matching (modal OAM), or phase matching (classical
OAM)? Is it possible to observe quantum OAM transfer in an optical
tweezer? Although we have indications that the integrand of eq. \eqref{eq:average_Lz}
is a classical probability of finding a given OAM value, a proper
verification could be given by obtaining angular velocity histograms
in an optical tweezer setup as \citep{DholakiaPerfectVortexDynamics2013,DholakiaOAMMathieubeams2006,RaizenTweezerinVacuum2011}.
Also, the proposed histograms are amenable to theoretical modelling,
that can be used to retrieve quantitatively the experimental parameters
of shaped OV beams and their uncertainties. A final remark is that
eq. \eqref{eq:average_Lz} allows a much simpler and faster way to
calculate the average OAM for a beam with known amplitude and phase
profile than decomposing it in a basis of OAM eigenmodes.

We acknowledge the financial support from the Brazilian agencies CNPq
(INCT-Fotônica) and FACEPE. We also acknowledge helpful discussions
with Dr. L. Pruvost. A. M. A. also thanks Dr. W. Löffler for inciting
an extension of our work on shaped OV \citep{Shaping_Optical_Beams}.

\bibliographystyle{7C__Users_Anderson_Dropbox_Pesquisa_artigos_Exp___OAM_in_complex_shaped_vortices_submissao_ol}
\bibliography{6C__Users_Anderson_Dropbox_Pesquisa_artigos_Exp___mplex_shaped_vortices_submissao_referencias}

\newpage{}

\section*{Supplementary informations}

We consider a complete basis of OAM eigenfunctions labelled by $p$
and $m$ as, respectively the radial and the azimuthal quantum numbers.
Using the completeness relation $1=\sum_{p,m}\left|p,m\left\rangle \right\langle p,m\right|$
it may be shown that the average OAM is 
\begin{equation}
\left\langle L_{z}\right\rangle =\left\langle \mathcal{A}\left|\hat{L}_{z}\right|\mathcal{A}\right\rangle =\sum_{p,m}m\left|\left\langle \mathcal{A}|p,m\right\rangle \right|^{2}.
\end{equation}

Meanwhile, one may express the average in terms of spatial wave functions
$\mathcal{A}\left(\mathbf{r}\right)=\left\langle \mathbf{r}|\mathcal{A}\right\rangle $.
\begin{align}
\left\langle L_{z}\right\rangle  & =\int d^{2}r\left\langle \mathcal{A}|\mathbf{r}\right\rangle \left\langle \mathbf{r}\left|\hat{L}_{z}\right|\mathcal{A}\right\rangle ,\\
 & =\int d^{2}r\left\langle \mathcal{A}|\mathbf{r}\right\rangle \hat{L}_{z}\left\langle \mathbf{r}|\mathcal{A}\right\rangle .\label{eq:avg_Lz_full_decomp}
\end{align}
Since $\left\langle L_{z}\right\rangle $ is real, the imaginary part
of the integrand in Eq. \eqref{eq:avg_Lz_full_decomp} must cancel.
Therefore the classical OAM density is given by
\begin{equation}
\left\langle L_{z}\right\rangle \left(\mathbf{r}\right)=\mathfrak{Re}\left\{ \left\langle \mathcal{A}|\mathbf{r}\right\rangle \hat{L}_{z}\left\langle \mathbf{r}|\mathcal{A}\right\rangle \right\} ,\label{eq:Lz_classical_from_quantum}
\end{equation}
or, by expanding $\left|\mathcal{E}\right\rangle $ in the OAM eigenfunctions,
\begin{equation}
\left\langle L_{z}\right\rangle \left(\mathbf{r}\right)=\mathfrak{Re}\left\{ \sum_{\begin{array}{c}
p,m\\
p',m'
\end{array}}m\left\langle \mathcal{A}|p,m\right\rangle \left\langle p',m'|\mathcal{A}\right\rangle \left\langle \mathbf{r}|p',m'\right\rangle \left\langle p,m|\mathbf{r}\right\rangle \right\} .\label{eq:Quantum_OAM_density}
\end{equation}

Equation \eqref{eq:Lz_classical_from_quantum} may also be expressed
in terms of the beam phase profile. Using that $\hat{L}_{z}=-i\frac{\partial}{\partial\phi}$
and expressing $\left\langle \mathbf{r}|\mathcal{A}\right\rangle =\left|\mathcal{A}\right|\exp\left(i\chi\right)$,
it is possible to show that 
\begin{equation}
\left\langle L_{z}\right\rangle \left(\mathbf{r}\right)=\left\langle \mathcal{A}|\mathbf{r}\right\rangle \frac{\partial\chi}{\partial\phi}\left\langle \mathbf{r}|\mathcal{A}\right\rangle .\label{eq:Classical_OAM_density}
\end{equation}

Equations \eqref{eq:Quantum_OAM_density} and \eqref{eq:Classical_OAM_density}
are, respectively, the expressions used to represent the classical
OAM density profile from the modal and classical data in Figs. 4-6.
\end{document}